\documentclass[aps,prl,nofootinbib,showpacs,floatfix,twocolumn]{revtex4}  % for review and submission
\usepackage{bm}
\usepackage{latexsym}
\usepackage{dcolumn}
\usepackage{amsfonts,amssymb,amsmath}
\usepackage{graphicx,epsfig}
\usepackage{psfrag}
\usepackage{subfigure}
\usepackage{feynmf}
\usepackage{rotating}
\usepackage{hyperref}
\usepackage{color}

\hypersetup{
    bookmarks=true,         % show bookmarks bar?
    unicode=false,          % non-Latin characters in Acrobat’s bookmarks
    pdftoolbar=true,        % show Acrobat’s toolbar?
    pdfmenubar=true,        % show Acrobat’s menu?
    pdffitwindow=false,     % window fit to page when opened
    pdfstartview={FitH},    % fits the width of the page to the window
    pdftitle={My title},    % title
    pdfauthor={Author},     % author
    pdfsubject={Subject},   % subject of the document
    pdfcreator={Creator},   % creator of the document
    pdfproducer={Producer}, % producer of the document
    pdfkeywords={keyword1} {key2} {key3}, % list of keywords
    pdfnewwindow=true,      % links in new window
    colorlinks=true,       % false: boxed links; true: colored links
    linkcolor=red,          % color of internal links
    citecolor=cyan,        % color of links to bibliography
    filecolor=magenta,      % color of file links
    urlcolor=green,           % color of external links
    linktocpage=true
}

\def\beq{\begin{equation}}
\def\efq{\end{equation}}
\def\br{\begin{eqnarray}}
\def\er{\end{eqnarray}}
\def\benu{\begin{enumerate}}
\def\efnu{\end{enumerate}}

\begin{document}

% Please keep new commands to a minimum, and use \newcommand not \def to avoid
% overwriting existing commands. Example:
%\newcommand{\pcm}{\,cm$^{-2}$}	% per cm-squared

%%%%%%%%%%%%%%%%%%%%%%%%%%%%%%%%%%%%%%%%%%%%%%%%%%

%%%%%%%%%%%%%%%%%%% TITLE PAGE %%%%%%%%%%%%%%%%%%%

% Title of the paper, and the short title which is used in the headers.
% Keep the title short and informative.
\title[Parameter discordance in Planck and low-redshift measurements]{Parameter discordance in Planck CMB and low-redshift measurements: projection in the primordial power spectrum}

\author{Dhiraj Kumar Hazra} \email{hazra@bo.infn.it}
\affiliation{Istituto Nazionale Di Fisica Nucleare, Sezione di Bologna,Viale Berti Pichat, 6/2, I-40127 Bologna, Italy; Osservatorio di Astrofisica e Scienza dello Spazio di Bologna/Istituto Nazionale di Astrofisica, via Gobetti 101, I-40129 Bologna, Italy}
\author{Arman Shafieloo} \email{shafieloo@kasi.re.kr}
\affiliation{Korea Astronomy and Space Science Institute, Daejeon 34055, Korea; \\
University of Science and Technology, Daejeon 34113, Korea}
\author{Tarun Souradeep} \email{tarun@iucaa.in}
\affiliation{Inter-University Centre for Astronomy and Astrophysics, Post Bag 4, Ganeshkhind, Pune 411 007, India}

\date{\today}

\begin{abstract}
We discuss the discordance between the estimated values of the cosmological parameters from Planck assuming the concordance $\Lambda$CDM model and 
low-redshift measurements. In particular, we consider the Hubble constant mismatch between Planck temperature constraint for the $\Lambda$CDM model 
and the Riess {\it et. al.}~\cite{0004-637X-855-2-136} local measurements as well as the discordance between the estimated value of $S_8$ from Planck 
and some weak lensing surveys such as Kilo Degree Survey (KiDS-450) and Dark Energy Survey (DES) observations. The discordance can come from a wide 
range of non-standard cosmological or astrophysical processes as well as from some particular systematics of the observations. 
%Radiative transport kernel includes the effects from these non-standard cosmological processes. 
In this paper, without considering any particular astrophysical process or extension to the standard model at the background level, we seek solely 
to project the effect of these differences in the values of the key cosmological parameters on to the shape of the primordial power spectrum (PPS). 
%Since the primordial power spectrum from Planck is broadly consistent with a nearly scale invariant form, it is easy to locate the possible differences in the power at different cosmological scales that can account for such changes. 
In order to realise this goal, we uncover the shape of the PPS by implementing the Modified Richardson-Lucy algorithm (MRL) that fits the Planck temperature data 
as acceptably as the case of the standard model of cosmology, but with a Hubble constant consistent with local measurements as well as improving the  consistency between
the derived $S_8$ and $\sigma_8$ parameters with estimations of the weak lensing surveys. 

% to being consistent to  using Modified Richardson-Lucy algorithm. We find that this shape of the PPS additionaly resolves the tension of 
%$\sigma_8$ normalization and matter density between Planck and  Kilo Degree Survey (KiDS-450) observations.
\end{abstract}

\maketitle
\section{Introduction}

An intriguing discrepancy between the estimated values of the cosmological parameters from Planck CMB measurements and some local observations has 
persisted over the last couple of years. While measurements of the Hubble constant from the local Universe is converging at $H_0\sim73~{\rm km/s/Mpc}$ with 
progressively narrower errorbars~\cite{2001ApJ...553...47F,2016ApJ...826...56R,0004-637X-855-2-136,2018ApJ...861..126R}, Planck CMB measurements 
consistently estimate $H_0=67.8 \pm 0.9~{\rm km/s/Mpc}$ 
for the case of the concordance $\Lambda$CDM model~\cite{Ade:2015xua,Aghanim:2018eyx}. On the other hand we are also confronted with a discrepancy between growth 
of structure measurements from weak lensing surveys such as KiDS~\cite{2017MNRAS.465.1454H}, CFHT~\cite{2017MNRAS.465.2033J}, DES~\cite{2017arXiv170801530D}, HSC~\cite{Hikage:2018qbn} and 
that deducted from Planck CMB measurements as reflected in the estimated values of $S_8=\sigma_8 \sqrt{\frac{\Omega_{0m}}{0.3}}$. While most of these constraints on the cosmological 
parameters from different surveys are in fact model dependent estimations and, in particular, assume the concordance $\Lambda$CDM model of the Universe, the persistence
of the discrepancy has made it imperative to investigate and reveal the nature of these disagreements. In fact, if these 
tensions are real and become more statistically significant with future observations (and not attributable to different systematics), we may be forced to seek
physics beyond the concordance model of cosmology that
effectively require exciting extensions or modifications beyond the current model~\cite{Sahni:2014ooa, Zhao:2017cud, Shafieloo:2018gin}. 
In this work we seek a modified form of the primordial spectrum deviating from its conventional minimal power-law form within 
the context of the concordance model that satisfies different astrophysical and cosmological observations and also alleviates the tensions in the
different estimations of key cosmological parameters. We show that it is indeed possible to project the discrepancies onto the form of the primordial power spectrum (PPS) 
and the reconstructed form of the PPS might potentially hint toward some specific physics in the early Universe. Nevertheless, the shape of the reconstructed form of the
PPS might prove to be a guide toward the identification of some systematics in the pipelines of the data reduction and analysis of Planck CMB data as well. First we describe 
the formalism used for our analysis and reconstruction, then we present our results and state our conclusion at the end.   
%%%%%%%%%%%%%%%%%%%%%%%%%%%%%%%%%%%%%%%%%%%%%%%%%%%%%%%%%%%%%%%%%%%%%%%%%%%%%%%

\section{Formalism}~\label{sec:formalism}
The primary goal behind this work is to reconstruct a form of the primordial power spectrum that results in the same CMB observables of the best fit 
standard $\Lambda$CDM model while also being consistent with substantially different combinations of the background cosmological parameters
to accommodate some discrepancies observed.
To achieve that we invoke strong priors on the key background parameters such as Hubble constant $H_0$, matter density $\Omega_{\rm m}$ and $\sigma_8$ 
consistent with low redshift observations and then reconstruct the form of the PPS that results in the same CMB temperature observables corresponding to
the best fit $\Lambda$CDM model. For the purpose of reconstruction, we use the modified Richardson-Lucy (MRL) formalism that has been 
introduced in~\cite{HSSWMAP9} and has been subsequently used in~\cite{HSSFF,HSSPlanck} for parameter estimation and  consistency check 
besides aiming at finding the features in the primordial power spectrum. For early works on the Richardson-Lucy algorithm and its 
variants, see~\cite{Richardson:72,Lucy:1974yx,Shafieloo:2003gf,Shafieloo:2006hs,Shafieloo:2007tk}. 

Planck baseline is defined using standard $\Lambda$CDM model with power law form of primordial power spectrum. In addition the 
sum of neutrino mass is fixed to be $0.06$eV and the effective number of relativistic species is fixed to be 3.046. We use Planck
baseline best-fit power spectrum to Planck 2015 TT and lowT data and we reconstruct the form of primordial power spectra that matches
the Planck best fit power spectrum while accommodating different and specifically chosen set of background cosmological parameters. Therefore, 
we use the smooth best fit angular power spectrum ($\Lambda$CDM model to Planck TT data) as an input data in our algorithm. As the
input power spectrum is always positive, we do not use the binned reconstruction part even when the signal-to-noise ratio falls~\cite{HSSWMAP9}. 

In our analysis we assume that the signatures of non-standard physics that may give rise to 
the discordance in estimated parameters between Planck CMB and other observations can be substantially captured
by the features in the reconstructed PPS. This assumption is motivated from some of our previous analyses. 
In~\cite{Shafieloo:2007tk,HSSFF} we demonstrated that allowing a free form of the PPS increases the degeneracy 
in the background cosmological parameters substantially. In~\cite{HSSPlanck}, we showed, when effects of CMB lensing is not
included in the analysis, MRL projects its partial imprints in the PPS as some specific form of features.

%In this work we use 1000 iterations in the MRL reconstruction to include all the features in the PPS that could have some effect. 
Since we are using theoretical best fit and not the data, we do not have the disadvantage of fitting noise. In particular, for 
Planck 2013 data combined from all frequency channels, we had demonstrated in~\cite{HSSPlanck} that with higher 
iterations, CMB lensing effect can be significantly mimicked by the reconstructed features.

To start with and to calculate the radiative transport kernel we fix $H_0=73.48 {~\rm km/s/Mpc}$~\cite{0004-637X-855-2-136}.
To be consistent with results from weak lensing surveys we consider the matter density to be lower than the derived 
values for the case of the standard $\Lambda$CDM model fitting the Planck data. We find that using the Planck TT + lowT best fit 
for $\omega_{\rm b}=\Omega_{\rm b}h^2$ and $\omega_{\rm CDM}=\Omega_{\rm CDM}h^2$ provides a value of $\Omega_{\rm m}=0.259$ if
$H_0=73.48 {~\rm km/s/Mpc}$ is used which is suitable for our purpose. While heights the CMB acoustic 
peaks and dips are sensitive to $\omega_{\rm b}$ and $\omega_{\rm CDM}$~\cite{Hu:2000ti, Aghamousa:2013jvj, Aghamousa:2014nca}, 
using the same values for our kernel results in minimal features in the form of the primordial spectrum at large scales. 
% We should mention again that the 
% Planck best-fit standard $\Lambda$CDM baseline angular power spectrum is used as the input data in our analysis. 
% Then we use our MRL algorithm to reconstruct the form of the primordial spectrum. At the end we use Planck likelihood code
% to test if indeed the combination of the background parameters and the reconstructed PPS can provide a likelihood comparable
% with the case of the best fit $\Lambda$CDM model.  

Once we have the reconstructed form of the PPS from our MRL algorithm, we include scope for some fine tuning (in order to maximize the likelihood)
by varying $A_{\rm scale}$ the overall amplitude of the reconstructed PPS, allowing a lateral shift of the feature positions using
$\Delta\ln k$ and also performing a Gaussian smoothing of the features with a smoothing width of $\Delta_{\rm smooth}$.  Next we 
use the reconstructed PPS for cosmological background parameter estimation. While in the case of the standard $\Lambda$CDM model 
one assumes a power-law form of the PPS to perform parameter estimation, here we fix the form of the PPS to be what we have reconstructed 
and then establish how much cosmological parameters can vary while satisfying Planck temperature observations. In particular we are 
interested to see how much $\Omega_{\rm m}$ and $H_0$ can vary around their fixed assumed values used to reconstruct the primordial power spectrum.

For background parameters we use the conventional parameters $\Omega_{\rm b}h^2$, $\Omega_{\rm CDM}h^2$, $\theta$ and $\tau$ that denotes baryon and cold dark matter densities, the ratio of the sound horizon to the angular diameter distance at decoupling and the Thomson scattering optical depth respectively. We use {\tt CAMB}~\cite{Lewis:1999bs} and {\tt CosmoMC}~\cite{Lewis:2002ah} for parameter 
estimation. We calculate the angular power spectra at all multipoles without interpolation. We use publicly available Planck 2015 data and likelihood codes~\cite{Aghanim:2015xee,Ade:2015xua}. In particular we 
use Planck TT + lowT data for comparison. We do not use the polarization data at this stage to follow a conservative approach.

One important issue here to note is the CMB lensing. In our reconstruction procedure and the likelihood analysis we considered the 
lensing effect for different forms of the PPS and the choice of the background parameters. 

In MRL formalism the angular power spectrum data is used in unlensed format (${\cal C}_{\ell}^{\rm unlensed}$) since the effect of lensing needs to be subtracted 
in order to connect the primordial spectrum to angular power spectrum via our kernel which is a function of the background parameters.
\begin{equation}
{\cal C}_{\ell}^{\rm lensed}={\cal C}_{\ell}^{\rm unlensed}+{\cal C}_{\ell}^{\rm lensing\hspace{1mm}template} 
\end{equation}

In our work we aim to have the same CMB temperature angular power spectrum observables of the best fit $\Lambda$CDM model for the case of our reconstruction. Hence to consider the lensing effect in our analysis, we should have the equality relation,

\begin{widetext}
\begin{equation}
{\cal C}_{\ell}^{\rm unlensed}[\rm Planck~best~fit]+{\cal C}_{\ell}^{\rm lensing\hspace{1mm}template}[\rm Planck~best~fit] ={\cal C}_{\ell}^{\rm unlensed}[\rm assumed\hspace{1mm}model]+{\cal C}_{\ell}^{\rm lensing\hspace{1mm}template}[\rm assumed\hspace{1mm}model]\end{equation}
\end{widetext}

Note that in our analysis the ${\cal C}_{\ell}^{\rm unlensed}[\rm assumed\hspace{1mm}model]$ is used for the reconstruction and our assumed model described
described earlier has the combination of $\Omega_{\rm m}=0.259$ and $H_0=73.48 {~\rm km/s/Mpc}$ where $\omega_{\rm b}$, $\omega_{\rm CDM}$ and $\tau$ are the 
same values of the best fit $\Lambda$CDM model. 

%%%%%%%%%%%%%%%%%%%%%%%%%%%%%%%%%%%%%%%%%%%%%%%%%%%%%%%%%%%%%%%%%%%%%%%%%%%%%%%
\section{Results}\label{sec:results}
\begin{figure}
\resizebox{250pt}{200pt}{\includegraphics{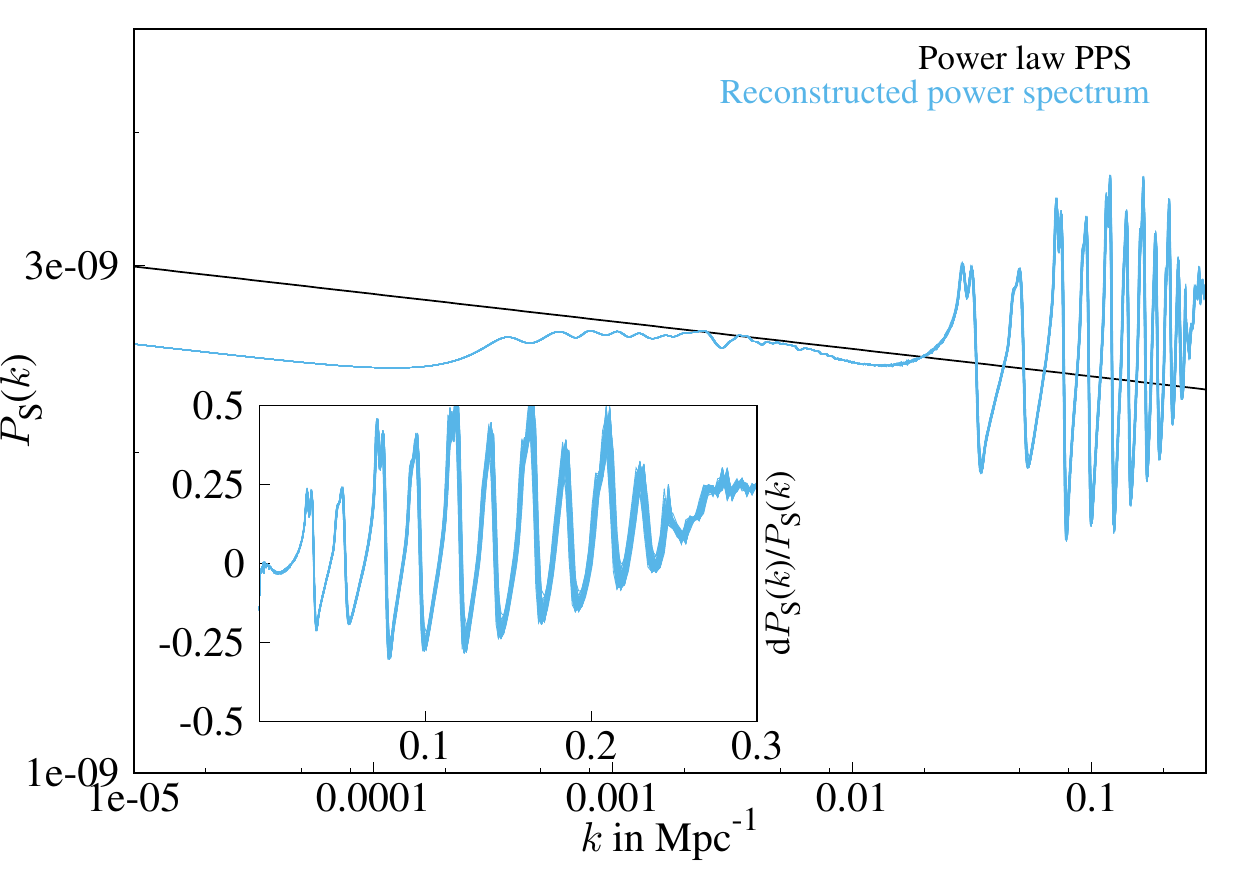}}
\caption{\label{fig:psk} 
The shape of the reconstructed primordial power spectra assuming $H_0=73.48 {~\rm km/s/Mpc}$ and $\Omega_{\rm m}=0.259$. 
While for this combination $\omega_{\rm b}$, $\omega_{\rm CDM}$ have the same values as the case of the best fit $\Lambda$CDM model, 
these parameters can satisfy local $H_0$ measurement as well as the constraints from weak lensing surveys. Different power spectra 
are drawn from the samples that are consistent with the data after lateral shifts and smoothing to the original reconstructed spectrum.
}

\end{figure}
\begin{figure}
\resizebox{240pt}{240pt}{\includegraphics{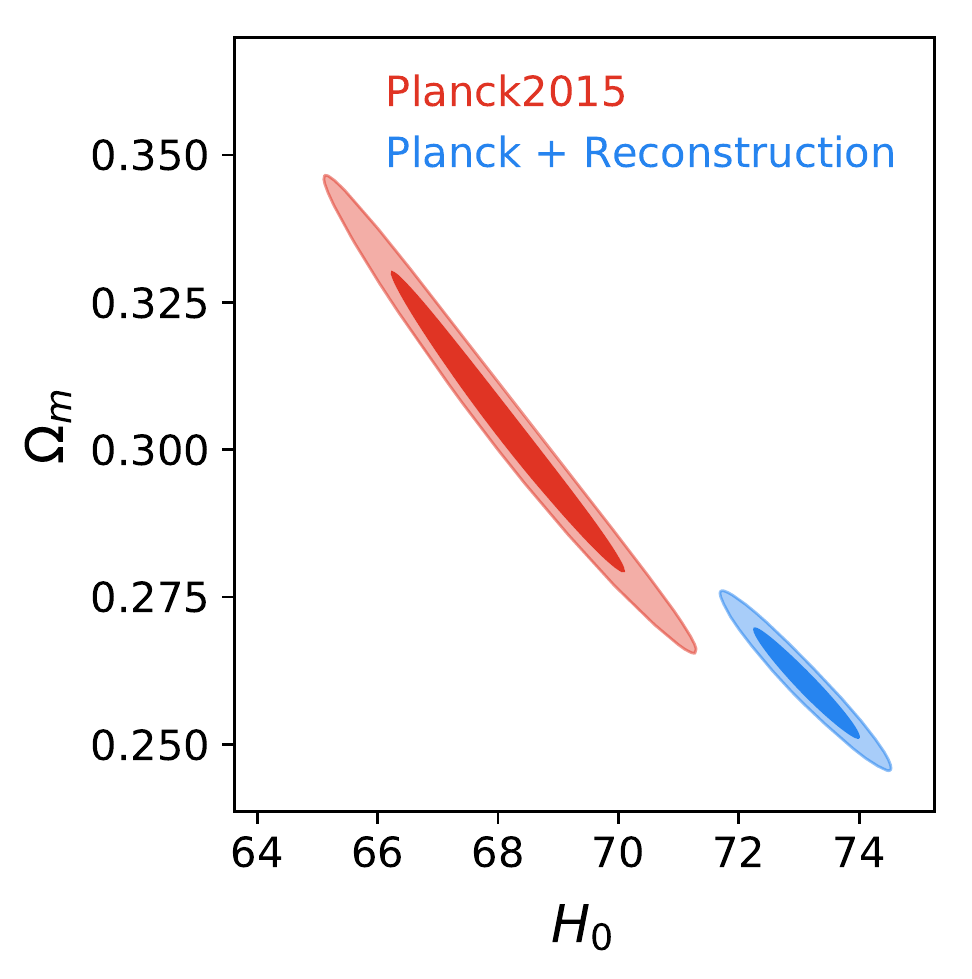}}
\caption{\label{fig:omm-h0} Marginalized 68\% and 95\% confidence contours in the $\Omega_{\rm m}$ and $H_0$ plane. Planck 2015 constraints using power law as PPS (in red) and using reconstructed PPS (in blue) are shown. Note that the reconstruction prefers a higher value of $H_0$ that matches with the local Hubble parameter measurement.}  
\end{figure}
\begin{figure}
\resizebox{240pt}{240pt}{\includegraphics{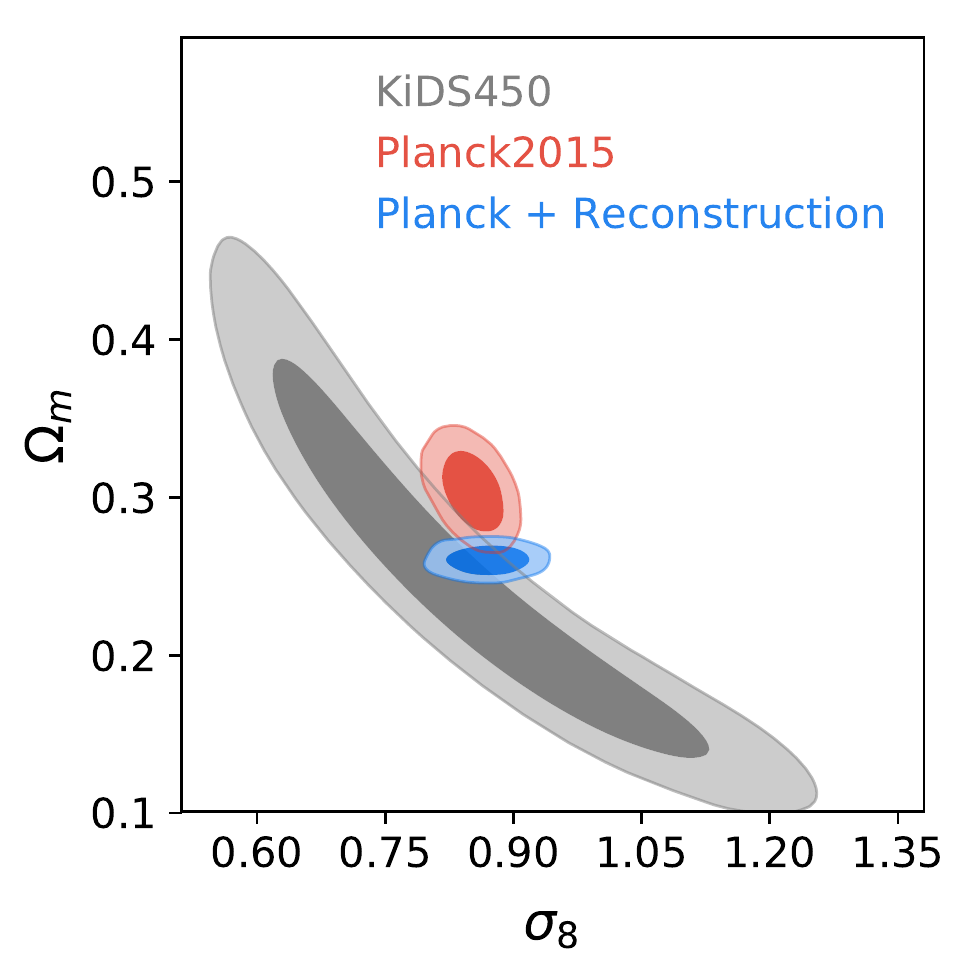}}
\caption{\label{fig:omm-s8} Marginalized 68\% and 95\% confidence contours of 
$\Omega_{\rm m}$ and $\sigma_8$ normalization. Planck 2015 constraints using power law PPS (in red) and using reconstructed PPS
(in blue) are shown while KiDS-450 weak lensing constraints are shown in grey. Note that the discrepancy is completely removed in the case of our reconstruction where there is an overlap between the 1$\sigma$ regions. %We find that the reconstruction that helps matching the Hubble parameter between Planck and local Hubble measurement, automatically resolves the matter density and normalization problem. However, the reconstructed PPS does require fine-tuned and detailed features.
}  
\end{figure}
\begin{figure*}
\resizebox{160pt}{120pt}{\includegraphics{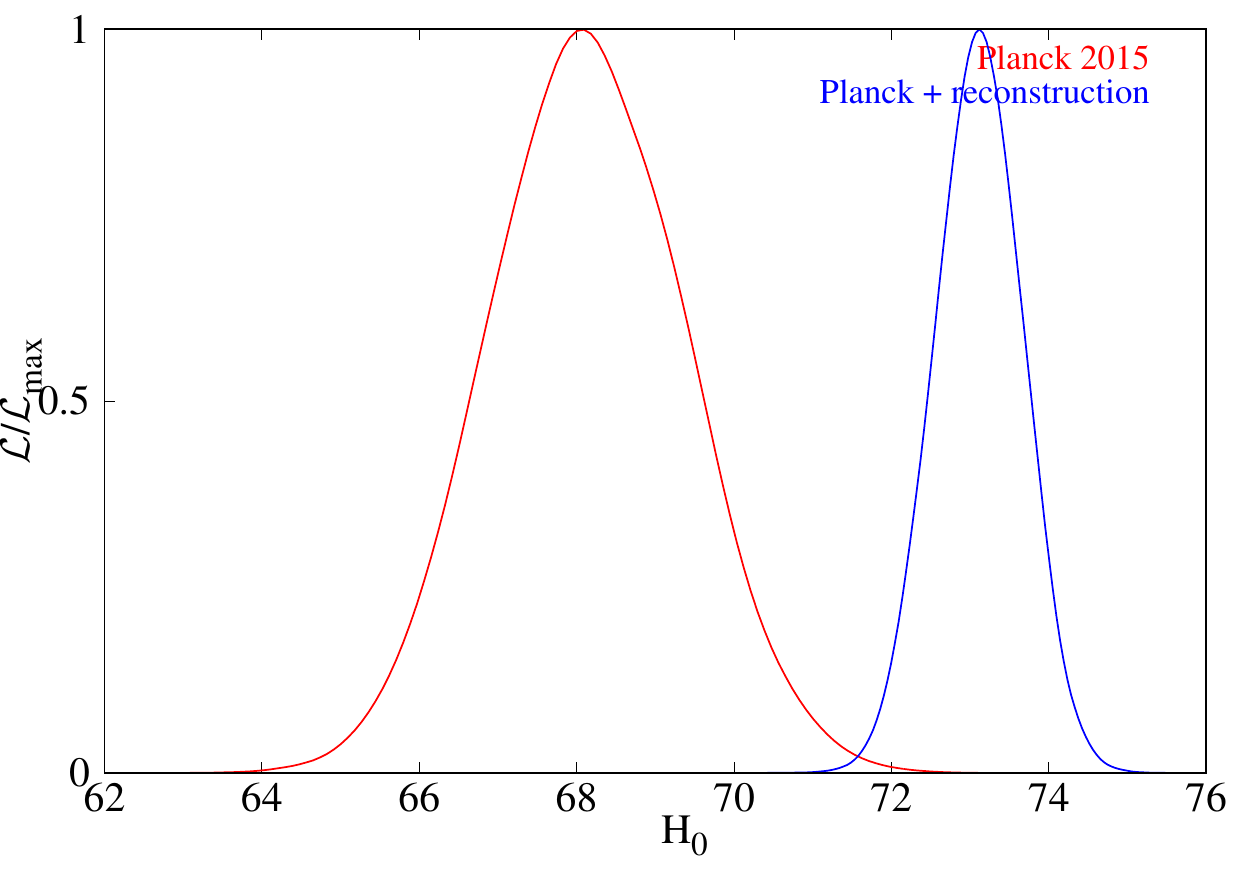}}
\resizebox{160pt}{120pt}{\includegraphics{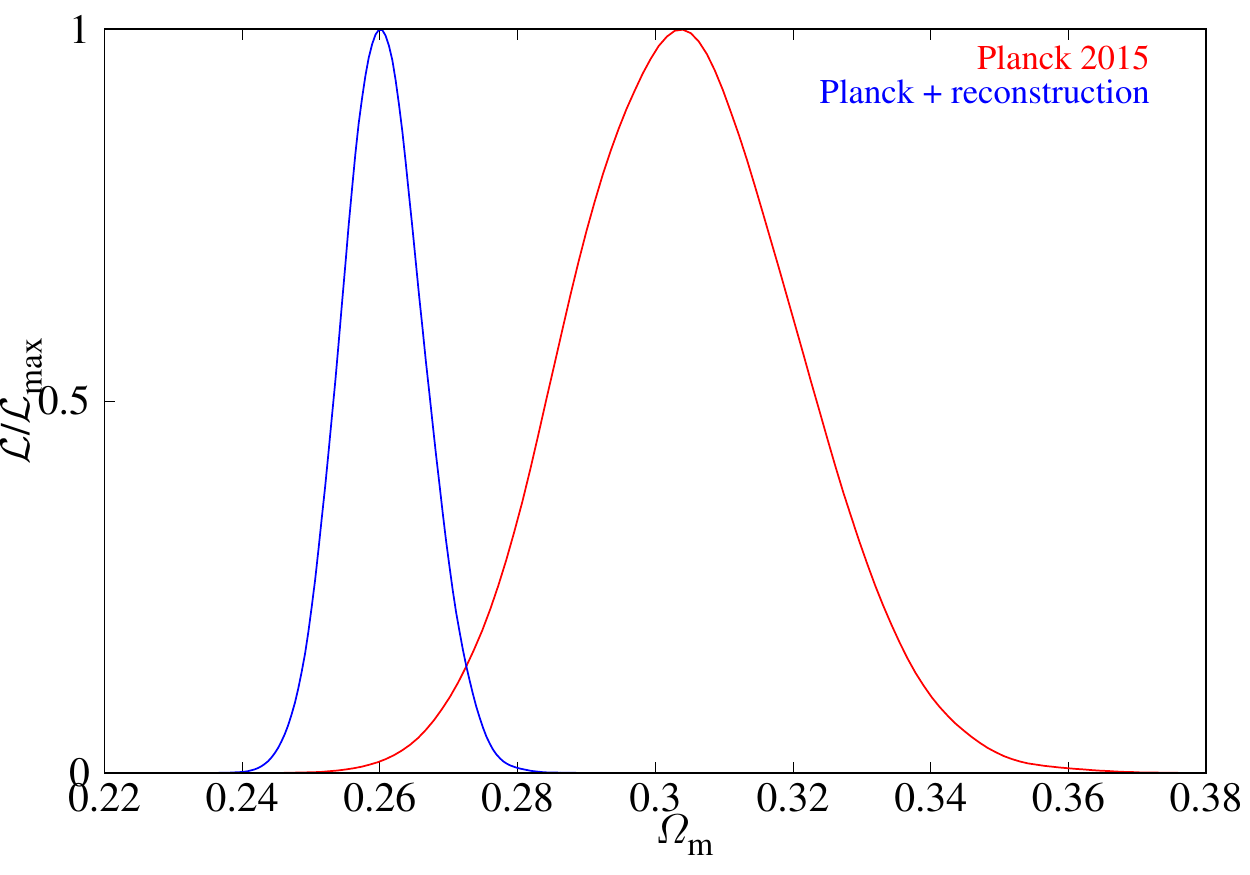}}
\resizebox{160pt}{120pt}{\includegraphics{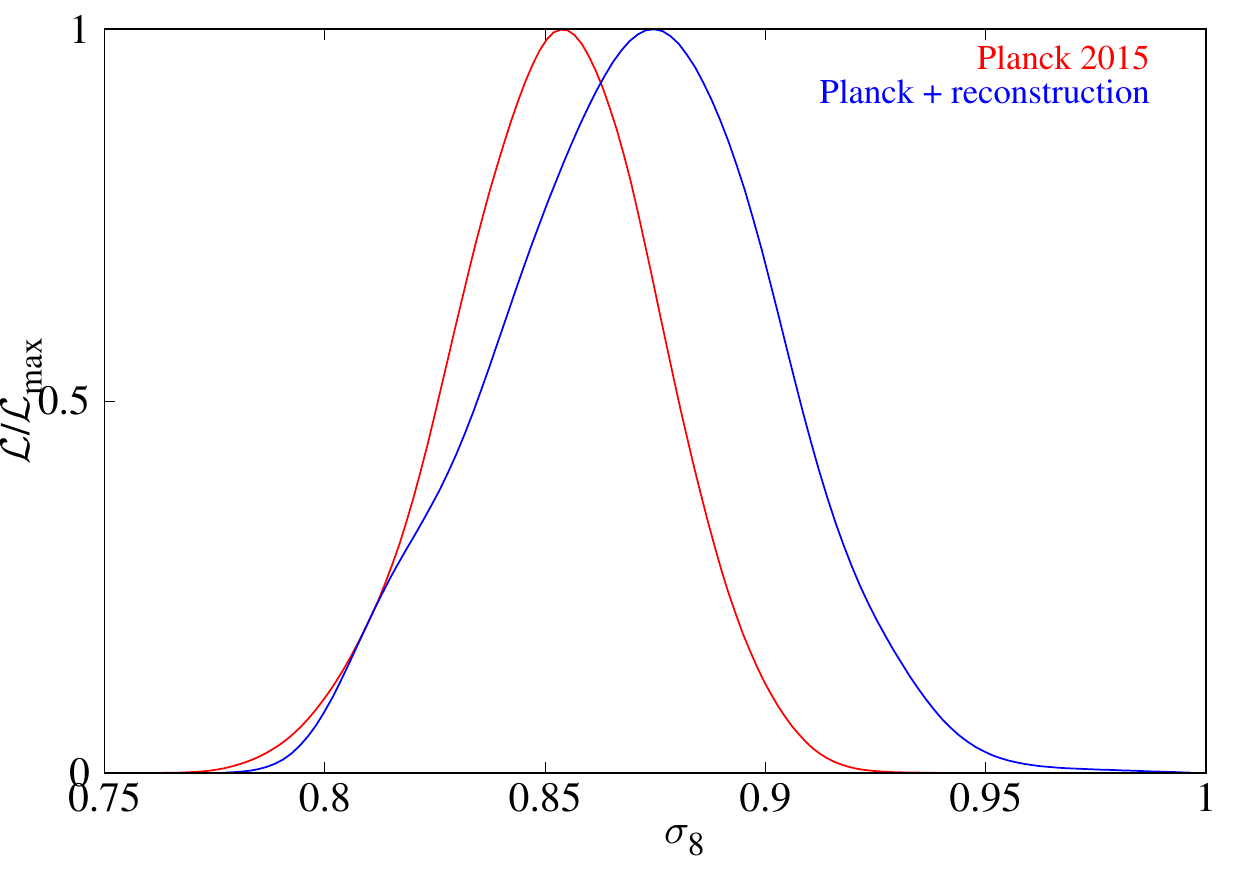}}
\caption{\label{fig:1Dandcorrelation} Marginalized probability distribution of $H_0$ [left], $\Omega_{\rm m}$ [centre], $\sigma_8$ [right] are plotted. The Hubble 
parameter constraints are shifted to higher values that agrees with the local measurement of $H_0$~\cite{0004-637X-855-2-136}. The mean value of matter density is substantially lower in the case of our reconstruction compared to the baseline case. Note that in both the cases the constraints are tighter in the case of reconstruction as its form has been fixed while there is a flexibility in the tilt of the power-law PPS. The $\sigma_8$ normalization constraints are comparable in both cases as we allow an overall amplitude shift of the PPS which is directly connected to the amplitude of the $\sigma_8$.}  
\end{figure*}

Fig.~\ref{fig:psk} shows the reconstructed forms of the PPS assuming $H_0=73.48 {~\rm km/s/Mpc}$ and $\Omega_{\rm m}=0.259$ where we also allow some 
lateral shifts or smoothing of the features after the reconstruction to maximize the likelihood. Results obtained overlap with
each other to generate a very narrow band. The reconstructed form of the PPS show some prominent features at different scales. While 
we see a suppression of power at large scales, we notice sharp fluctuations in the form of the PPS at wavenumbers larger than 0.02 $\rm Mpc^{-1}$.

We present a comparison of the parameter constraints between the standard $\Lambda$CDM model with power law PPS and the case of our study in Figs.~\ref{fig:omm-h0}
and~\ref{fig:omm-s8}. The comparison of the marginalized probabilities of these three parameters are provided in Fig.~\ref{fig:1Dandcorrelation}. 
The power law case is marked with Planck2015. Using the reconstructed PPS leads to high $H_0$ and low matter 
density around the assumed initial values (where we fixed the kernel for reconstruction) as expected. Using the reconstructed form of the PPS reduces the size of the confidence ball considerably because of the fine 
tuned features and no flexibility in the tilt of the reconstructed PPS. 

Constraints on $\sigma_8$ however remains similar, and even slightly larger than the case of the standard model with power-law form of the PPS as we allow the overall 
amplitude of the power spectrum to vary and $\sigma_8$ mainly depends on this  variation. We also plot the marginalized constraints on $\Omega_{\rm m}$ and $\sigma_8$ 
in Fig.~\ref{fig:omm-s8} from the KiDS observation~\cite{Joudaki:2016kym}
in the background. It is evident that the 3$\sigma$ tension can be removed by our specific form of PPS from reconstruction. In fact the marginalized estimated value of 
$S_8=0.811\pm0.03$ from our reconstruction fitting Planck CMB data is clearly more consistent with $S_8=0.745\pm0.038$ from KiDS,  $S_8=0.783^{+0.025}_{-0.021}$ from DES and $S_8=0.737\pm0.038$ 
from CFHTLenS (in comparison with the case of the power law PPS). It is intriguing that our reconstructed form of the PPS combined with the choice of the parameters we assumed, can indeed satisfy Planck CMB data as well
as the local Hubble measurement and also resolves the tension between the Planck measured values of matter density and matter power normalization $\sigma_8$ with that inferred using the lensing data from KiDS-450, DES and other weak lensing surveys. 

%attempt to fit the local Hubble measurement additionaly resolves the tension between the Planck measured values of matter density and matter power normalization $\sigma_8$ 
%with that inferred using the lensing data from KiDS-450.

\section{Discussion}\label{sec:discussion}
In this paper we project the inconsistencies in the estimated values of the key 
cosmological parameters assuming the concordance cosmological model from 
different surveys, to the form of the primordial spectrum. In other words, 
we reconstruct a form of the primordial spectrum that for a 
particular set of cosmological parameters (that are consistent with 
different low redshift observations), can result identically as the case
of the concordance $\Lambda$CDM model with power-law form of the PPS 
fitting Planck CMB temperature data. In the result section we show that 
this is possible. The particular form of the reconstructed PPS derived by enforcing 
the higher values of $H_0$ from local measurements also satisfies 
the lower value of $S_8$ estimated by various weak lensing surveys. While
the main purpose of this paper has been to demonstrate this possibility, 
there are few important issues to note: 

\begin{enumerate}
 \item The features at high $k$ values are very similar to the features we 
 reconstructed previously in~\cite{HSSPlanck} when we did not consider 
 CMB lensing (trying to project the effect on the form of the PPS). While 
 in our analysis we have considered the lensing effect for each point in 
 the parameter space, our results appear to suggest that the CMB lensing 
 templates prefer a background cosmology very close to the concordance 
 model with its best fit parameters. It is not so straightforward to 
 interpret this observation however, it might be worthy of further 
 investigation. Our results indirectly suggest that the lensing 
 templates which we use to analyze CMB data can have substantial 
 effect on the constraints on the form of the PPS and background 
 cosmological parameters. This might not be evident when we use a 
 power-law form of the PPS, but the issue brought to light more clearly when
 features in the form of the primordial spectrum are allowed. 

 \item At face value it appears to be unnatural to generate the complex
 form of the reconstructed PPS within an 
 inflationary scenario without extreme fine 
tuning. However, we do not provide any conclusive reason to close the
possibility of a physical early Universe explanation. In fact the
reconstructed form of the PPS at large scales might provide a hint
for a specific phenomenological model while at the small scales we 
might have to consider other issues such as
the lensing effects indicated in the earlier point. 

 \item Using polarization data it should be possible to validate further the 
 possibility of the reconstructed form of the PPS. Likewise, using polarization 
 data we might be able to look for a more optimized form of the PPS to remove 
 tensions from different observations. This is deferred to future study. 

 \item A wider exploration of the underlying parameter space of the cosmological
 model would be essential to reveal potential routes to ameliorate the disagreements 
 in cosmological parameters inferred. In this work we fixed two of the key cosmological parameters,
 $H_0$ and $\Omega_{\rm m}$ for reconstruction. An interesting possible 
 extension of this work where the parameters are held fixed for the MRL
 reconstruction, is then to allow them to vary within a particular range of interest. This is planned for future works. 
 
 \item We considered a combination of key cosmological parameters in order to 
 satisfy the local $H_0$ measurement as well as estimations of the $\sigma_8$ and 
 $S_8$ from weak lensing surveys. However, these constraints from weak lensing surveys
 are generally based on assumption of the concordance model itself where power-law 
 form of the primordial spectrum is assumed. While we have a different form of the 
 PPS from our reconstruction, observational constraints of $\sigma_8$ and $S_8$ might become 
 slightly different from what they have been reported! In fact a comprehensive analysis requires an iterative approach
 considering all the theoretical effects as well as the effects on observational 
 constraints. This is also beyond the scope of this work which requires a more
 generalized pipeline to estimate cosmological parameters. 
 
 \item Though we did not discuss in the paper, our result may also alleviate the tension 
 between the cosmological constant dark energy and the Ly-$\alpha$ forest BAO measurement 
 at $z=2.33$~\cite{Bautista:2017zgn}. The tension has been reported as a serious problem 
 for the cosmological constant and a hint for an evolving dark energy~\cite{Sahni:2014ooa,Zhao:2017cud}.
 By allowing a lower matter density (with respect to the standard model cosmology fitting Planck data), 
 the expansion history $h(z)$ in the case of $\Lambda$ dark energy can have substantially lower values
 at redshifts higher than two that may help to make this model more consistent with the Ly-$\alpha$ BAO observations.  
 
\end{enumerate}

\section*{Acknowledgments}
The authors would like to thank Shahab Joudaki for discussions regarding the KiDS results.
The authors would like to acknowledge the use of APC cluster (\href{https://www.apc.univ-
paris7.fr/FACeWiki/pmwiki.php?n=Apc-cluster.Apc-cluster}{https://www.apc.univ-
paris7.fr/FACeWiki/pmwiki.php?n=Apc-cluster.Apc-cluster}). A.S. would like to acknowledge the support 
of the National Research Foundation of Korea (NRF-2016R1C1B2016478).
\bibliography{HST-recon}

\end{document}